\documentclass[epj]{svjour}
\usepackage{graphicx,latexsym,amssymb,amsmath,color,multirow,mathrsfs,booktabs,ifpdf}
\usepackage{dcolumn}
\usepackage{bm}

\usepackage{float, varioref}
\usepackage{longtable}

\def\oc{$^{16}$O+$^{12}$C\ }

\def\cc{$^{12}$C+$^{12}$C\ }
\def\oo{$^{16}$O+$^{16}$O\ }
\def\AA{nucleus-nucleus\ }
\def\Atrans{$^{12}$C ($^{16}$O,$^{12}$C)$^{16}$O\ }
\begin{document}
\title{Elastic $\alpha$ transfer in the \oc scattering and its impact on the nuclear 
 rainbow}
\author{Nguyen Hoang Phuc\inst{1}\and Dao T. Khoa\inst{1}\and Nguyen Tri Toan Phuc\inst{1,2}} 
\institute{Institute for Nuclear Science and Technology, VINATOM, Hanoi 100000, Vietnam
\and University of Science, Ho Chi Minh City, Vietnam}
\date{Received: date / Revised version: date}

\abstract{Elastic \oc scattering is known to exhibit the nuclear rainbow pattern 
at incident energies $E_\text{lab}\gtrsim 200$ MeV, with the Airy structure 
of the far-side scattering cross section clearly seen at medium and large angles. 
Such a rainbow pattern is well described by the deep real optical potential (OP) 
given by the double-folding model (DFM). At lower energies, the extensive elastic 
\oc scattering data show consistently that the nuclear rainbow pattern 
at backward angles is deteriorated by an oscillating enhancement of elastic 
cross section that is difficult to describe in the conventional optical model 
(OM). Given a significant $\alpha$ spectroscopic factor predicted for the 
dissociation $^{16}$O$\to\alpha+^{12}$C by the shell model and $\alpha$-cluster 
models, the contribution of the elastic $\alpha$ transfer (or the core-core 
exchange) to the elastic \oc scattering should not be negligible and is expected 
to account for the enhanced elastic cross section at backward angles. 
To reveal the impact of the elastic $\alpha$ transfer, a systematic coupled 
reaction channels analysis of the elastic \oc scattering has been performed, 
with the coupling between the elastic scattering and elastic $\alpha$ transfer 
channels treated explicitly, using the real OP given by the DFM. We found that 
the elastic $\alpha$ transfer enhances the near-side scattering significantly 
at backward angles, giving rise to an oscillating distortion of the smooth 
Airy structure. The dynamic polarization of the OP by the coupling between 
the elastic scattering and elastic $\alpha$ transfer channels can be effectively 
taken into account in the OM calculation by an angular-momentum 
(or parity) dependent potential added to the imaginary OP, as suggested by 
Frahn and Hussein 40 years ago.} 
\maketitle

\section{Introduction}
Although elastic heavy-ion (HI) scattering is usually dominated by the strong 
absorption \cite{Sa79,Bra97}, some light HI and $\alpha$-nucleus systems are 
quite weakly absorbing so that the nuclear rainbow pattern survives at medium and 
large scattering angles. The observation of the nuclear rainbow allows the determination
of the real \AA optical potential (OP) with much less ambiguity (see the topical 
review \cite{Kho07r} for more details). The nuclear rainbow originates from the 
far-side scattering \cite{Hus84}, and usually is associated with a broad Airy 
oscillation \cite{Kho07r,Bra96}. The observation 
of the Airy minima, in particular, the first Airy minimum A1 that is followed by 
a shoulder-like bump is essential for the identification of the nuclear rainbow 
\cite{Kho07r,Bra96,Fri88}. The far-side scattering pattern of the nuclear rainbow 
can be revealed by the decomposition of the elastic scattering amplitude into the 
\emph{internal} component that penetrates the Coulomb + centrifugal barrier 
well into the interior of the real OP, and the \emph{barrier} component that is 
reflected from the barrier \cite{Bri77,Row77}. An alternative interpretation 
of the far-side scattering is the decomposition of the elastic scattering amplitude 
into the \emph{near-side} and \emph{far-side} components \cite{Ful75} which is referred 
to and discussed throughout this work. In any case, the observation of the nuclear
rainbow pattern provides very important database for the mean-field study of the 
refractive \AA scattering.  

The recent folding model analysis \cite{Kho16} of elastic \cc and \oc scattering
has pointed out a range of refractive energies ($10\lesssim E \lesssim 40$ MeV/nucleon 
for the incident $^{12}$C and $^{16}$O ions), where the nuclear rainbow pattern 
can be clearly identified. At lower energies, the rainbow shoulders following 
the Airy minima are located closer to backward angles, and are deteriorated 
by the Mott interference in the symmetric \cc and \oo systems or by the elastic 
$\alpha$ transfer in the \oc system. On the other hand, the broad Airy structure 
is moving to forward angles with the energies increasing above the refractive range, 
and the nuclear rainbow pattern is destroyed by the interference of the near-side 
and far-side scatterings that leads to the well-known Fraunhofer oscillation. 

The elastic \cc and \oo scattering is strongly refractive and favorable for 
the observation of nuclear rainbow, like the pronounced primary rainbow observed 
in the elastic \oo scattering at $E_\text{lab}=350$ MeV \cite{Sti89}. However, 
the Airy structure of these systems is destroyed at scattering angles 
$\theta_\text{c.m.}\gtrsim 90^\circ$ because the boson symmetry of two identical 
nuclei gives rise to a rapidly oscillating elastic cross section there. For this 
reason, the asymmetric $^{16}$O+$^{12}$C system was considered as a good candidate 
for the study of nuclear rainbow \cite{Bra91}. 
Numerous experiments have been performed so far to measure the elastic \oc 
scattering with high-precision, over a wide range of energies 
($E_\text{lab}\approx 20-1503$ MeV) and a broad angular region 
(up to $\theta_\text{c.m.}>130^\circ$ at low energies) 
\cite{Rou85,Bra86,Vil89,Oglo98,Oglo00,Glu01-181,Glu07,Nico00,Bra01}.
Very essential for the present study are the elastic \oc data measured 
by the Kurchatov group \cite{Oglo98,Oglo00,Glu01-181,Glu07}, which exhibit the
pronounced Airy structure of nuclear rainbow \cite{Kho16,Oglo00}, and the  
data measured at lower energies by the Strasbourg group \cite{Nico00}. The  
optical model (OM) and folding model studies of the elastic data measured for the 
\cc, \oo, and \oc systems \cite{Kho16,Kho94,Kho97,Bra88,Kho00a,Mich01,Ohku14-1} 
have shown unambiguously the nuclear rainbow pattern, especially, the evolution 
of the Airy structure with the energy. However, the elastic \oc data measured 
at low energies ($E_\text{lab}\lesssim 132$ MeV) show that the nuclear rainbow 
pattern is substantially deteriorated by a quick oscillation of elastic cross 
section at backward angles.

\begin{figure}[bht]\vspace*{0cm}
\includegraphics[width=0.5\textwidth]{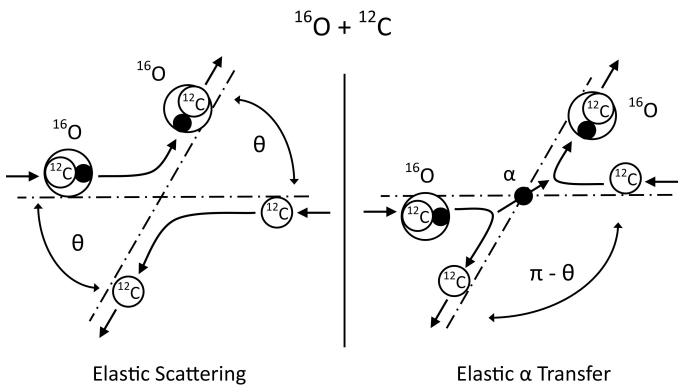}\vspace*{0cm}
 \caption{Kinematical illustration of the elastic scattering and elastic $\alpha$ 
 transfer processes in the \oc system.} \label{f1}
\end{figure}
The enhanced oscillatory cross section at backward angles, known as the ``anomalous
large angle scattering'' (ALAS) was observed for different light HI systems at 
low energies \cite{Brau82}. All these data show consistently that the surface 
of the dinuclear system is more transparent and the oscillating elastic 
cross section at backward angles is resulted from an interference process 
that can be generated in the OM calculation using an explicit angular-momentum 
(or parity) dependent term of the OP \cite{Brau82}. Such a procedure was shown 
by Frahn and Hussein \cite{Frahn80,Frahn80s,Frahn84} to result on the modified 
elastic $S$-matrix that contains a parity dependent component. While the oscillation 
of the low-energy elastic \cc or \oo cross section at backward angles is caused 
mainly by the boson symmetry of two identical nuclei \cite{Kho16,Kho00a,Mich01}, 
the direct and indirect $\alpha$ transfer was shown recently as the main 
physics origin of the oscillating enhancement of the elastic \oc cross section 
at backward angles \cite{Phuc18}. In general, the OM analysis of low-energy 
elastic \oc data is more difficult compared to other light HI systems 
\cite{Brau82}, and an $\ell$-dependent term was often added to the complex OP, 
which was suggested by von Oertzen and Bohlen \cite{vOe75} as necessary to account 
for the core-core exchange or elastic $\alpha$ transfer between $^{16}$O and 
$^{12}$C (see Fig.~\ref{f1}). Within the OM using standard parity-independent 
Woods-Saxon (WS) potential, one could obtain a good description of the elastic 
\oc data at low energies only if an extremely small diffuseness of the imaginary 
Woods-Saxon (WS) potential is used \cite{Nico00}. Such an abrupt shape of the 
absorptive WS potential is drastically different from the global OP established 
for the \oc system \cite{Bra97}.

The direct $\alpha$ transfer in the elastic $^{16}$O+$^{12}$C scattering at low 
energies were studied in the distorted wave Born approximation (DWBA) 
\cite{Mor11,Hama11}, and a good DWBA description of elastic \oc data was obtained 
with the $\alpha$ transfer amplitude added to the elastic scattering amplitude. 
Some scenarios of the $\alpha$ transfer in elastic \oc scattering were considered 
within the coupled reaction channel (CRC) formalism \cite{Szi02,Rud10}, which is
the most appropriate method to study transfer reactions. A detailed CRC analysis 
of the $\alpha$ transfer in the elastic \oc scattering at low energies was done
recently \cite{Phuc18}, where the multistep coupling of the elastic scattering channel
to the inelastic scattering, direct and indirect $\alpha$ transfer channels was treated 
explicitly, and a good CRC description of the data was obtained using the $\alpha$ 
spectroscopic factors predicted by the large-scale 
shell-model (SM) calculation \cite{Volya15}.

Given the $\alpha$ transfer in the elastic \oc scattering now well established, 
we focus in the present paper on how the $\alpha$ transfer affects the nuclear 
rainbow pattern in the elastic \oc cross section at large scattering angles. 
          
\section{Optical model description of the elastic \oc scattering and nuclear rainbow}
\label{sec2}
\begin{figure}\vspace*{-0.3cm}\hspace*{-1.0cm}
\includegraphics[width=0.55\textwidth]{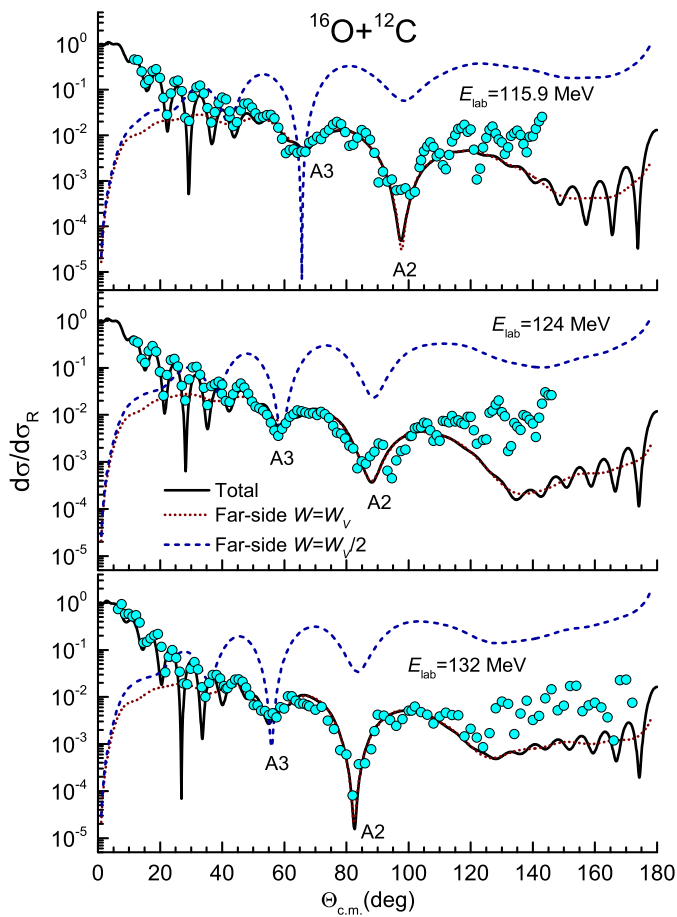}\vspace*{-1.5cm}
 \caption{OM description of the elastic \oc data measured at 
$E_{\rm lab}=115.9$, 124 MeV \cite{Nico00}, and 132 MeV \cite{Oglo98,Oglo00} using 
the real folded and WS imaginary OP (solid lines). The far-side cross sections 
are given by the near-far decomposition (\ref{eq1}) using the same real OP but 
with different absorptive strengths $W_V$ of the WS imaginary OP (dashed and 
dotted lines). Ak is the k-th order Airy minimum.} \label{f2}
\end{figure}

A realistic choice of the complex OP for elastic \oc scattering is a prerequisite 
for both the identification of nuclear rainbow and CRC study of the nonelastic 
reaction channels. In particular, the Fraunhofer oscillation at forward angles is 
formed entirely by elastic scattering, and a properly chosen OP for the \oc system 
should reproduce the elastic data at small angles as accurately as possible. 
With the energy increasing to about 10 to 40 MeV/nucleon, the nuclear rainbow pattern 
becomes well observable, and the most pronounced rainbow pattern associated with 
the first Airy minimum A1 was identified in the elastic \oc data measured at 
$E_{\rm lab}=200$ MeV \cite{Oglo00}. Such data were often used to validate different 
theoretical models of the \AA OP, like the double-folding model (DFM) that calculates 
the real OP using the realistic wave functions of two colliding nuclei and 
effective nucleon-nucleon interaction between the projectile- and target nucleons 
\cite{Kho16,Kho94,Kho97}. In the present work, the real OP given by the DFM is used in both 
the OM and CRC calculations. At low energies the imaginary OP is due to a few 
nonelastic channels and is parametrized in the standard WS form. The details of the 
DFM calculation of the real OP, Coulomb potential, and the WS parameters of the 
imaginary OP are given in Ref.~\cite{Phuc18}. The obtained OM description of the 
elastic \oc scattering data measured at incident energies of 115.9, 124 MeV 
\cite{Nico00}, and 132 MeV \cite{Oglo98,Oglo00} is shown in Fig.~\ref{f2}.  

To show the Airy structure of the nuclear rainbow pattern we decomposed 
the elastic scattering amplitude into the near-side and far-side components 
using Fuller's method \cite{Ful75}. Namely, by splitting the Legendre function 
$P_\ell(\cos\theta)$ into two waves scattered at $\theta$ but running in the
opposite directions around the scattering center, the elastic scattering 
amplitude $f_{\rm ES}(\theta)$ can be expressed in terms of the near-side 
($f_{\rm N}$) and far-side ($f_{\rm F}$) components as
\begin{eqnarray}
  f_{\rm ES}(\theta)&=&f_{\rm N}(\theta)+f_{\rm F}(\theta)=\frac{i}{2k}
	\sum_\ell (2\ell+1)A_\ell \nonumber\\
	&\times&\left[\tilde Q_\ell^{(-)}(\cos\theta)+
	\tilde Q_\ell^{(+)}(\cos\theta)\right], \label{eq1}
 \end{eqnarray}
 \begin{equation}
 {\rm where}\ \ \tilde Q_\ell^{(\mp)}(\cos\theta)={1\over 2}
  \left[P_\ell(\cos\theta)\pm {2i\over\pi}Q_\ell(\cos\theta)\right], \nonumber
  \end{equation}
and $Q_\ell(\cos\theta)$ is the Legendre function of the second kind. 
$f_{\rm N}(\theta)$ represents the wave deflected to the direction of $\theta$ 
on the near side of the scattering center, and $f_{\rm F}(\theta)$ represents 
the wave traveling on the opposite, far side of the scattering center to the 
same angle $\theta$. Thus, $f_{\rm N}(\theta)$ accounts mainly for the 
diffractive scattering that occurs at the surface, and $f_{\rm F}(\theta)$ 
accounts for the refractive scattering that penetrates more into the interior 
of the dinuclear system. The far-side scattering cross sections given by the 
best OM fit to the elastic \oc data measured at $E_{\rm lab}=115.9$, 124, and
132 MeV are shown as dotted lines in Figs.~\ref{f2}, and the broad rainbow 
shoulders following the second (A2) and third (A3) Airy minima are clearly 
seen. At these low energies, the first Airy minimum A1 is rather weak and
located at backward angles (on the ``dark'' side of rainbow). Because the 
refractive Airy structure of the far-side scattering is frequently 
obscured by the absorption, the OM calculation was done also with a strength 
of the imaginary WS potential reduced by 50\%, and the far-side cross sections 
(see the dashed lines in Fig.~\ref{f2}) show clearly the broad Airy oscillation 
pattern of nuclear rainbow which was established earlier in the extensive OM 
analyses of elastic \oc data over a wide range of energies \cite{Kho16,Oglo00}. 
All the OM calculations were done using the code ECIS97 written by Raynal \cite{Raynal}.

\section{CRC description of the elastic alpha transfer 
 $^{12}$C ($^{16}$O,$^{12}$C)$^{16}$O} \label{sec3}
Given the core-core symmetry of the \oc system, the elastic $\alpha$ transfer from
$^{16}$O to $^{12}$C leads to the final state that is indistinguishable from that 
of pure elastic scattering (see Fig.~\ref{f1}). Therefore, the total elastic amplitude 
must be a coherent sum of the elastic scattering (ES) amplitude $f_{\rm ES}$ and elastic 
transfer (ET) amplitude $f_{\rm ET}$. The interference between $f_{\rm ES}$ and 
$f_{\rm ET}$ was shown by the recent multichannel CRC analysis of elastic \oc 
scattering \cite{Phuc18} to give rise to an enhanced oscillating elastic cross 
section at large angles, using the $\alpha$ spectroscopic factors predicted by the
large-scale SM calculation \cite{Volya15} and 4$\alpha$ cluster model of $^{16}$O
\cite{Yama12}. Because the multichannel coupling of the indirect reaction channels
to the elastic \oc scattering channel can be accounted for by using an effective 
$\alpha$ spectroscopic factor adjusted to the best CRC fit to elastic data, taking 
into account the direct $\alpha$ transfer only \cite{Phuc18}, it is 
sufficient to focus on the coupling between the elastic scattering 
$^{12}$C ($^{16}$O,$^{16}$O)$^{12}$C and direct $\alpha$ transfer \Atrans channels
in the present study. Thus, the CRC equations for the two considered channels are 
solved using the code Fresco written by Thompson \cite{Tho88}
\begin{align}
 (E_a-T_a-U_a)\chi_a = [\langle a|W_b|b\rangle +
 \langle a|b\rangle(T_b+U_b-E_b)]\chi_b \nonumber \\ 
 (E_b-T_b-U_b)\chi_b = [\langle b|W_a|a\rangle +
 \langle b|a\rangle(T_a+U_a-E_a)]\chi_a \nonumber
\end{align} 
where $|a\rangle\equiv|^{16}$O,$^{12}$C$\rangle$ and $|b\rangle\equiv|^{12}$C,$^{16}$O$\rangle$, 
with the interchange of projectile and target; $\chi_a$ and $\chi_b$ are the scattering
wave functions of the elastic scattering and $\alpha$ transfer channels, respectively, and  
$U_a$ and $U_b$ are the corresponding OP's. Due to the identity of two channels, all post 
form formulas are equivalent to the prior ones and transfer interaction is determined 
\cite{Sat83,Tho09} as 
\begin{equation}
 W_a=W_b= V_{\alpha+^{12}\text{C}}+(U_{^{12}\text{C}+^{12}\text{C}}-
 U_{^{16}\text{O}+^{12}\text{C}}), \nonumber
\end{equation}
where $(U_{^{12}\text{C}+^{12}\text{C}}-U_{^{16}\text{O}+^{12}\text{C}})$ 
is the remnant term and $V_{\alpha+^{12}\text{C}}$ is the binding potential of
the $\alpha$ cluster in $^{16}$O. Further details of the CRC calculation 
can be found in Ref.~\cite{Phuc18}. An important input for the CRC calculation 
is the dinuclear overlap $\langle{\rm ^{12}C}|^{16}{\rm O}\rangle$ that is 
directly proportional to the $\alpha$ spectroscopic factor $S_\alpha$. 
The solutions $\chi_a$ and $\chi_b$ of the CRC equations are used to determine 
$f_{\rm ES}$ and $f_{\rm ET}$, and the total elastic \oc cross section is 
obtained as  
\begin{equation}
\frac{d\sigma(\theta)}{d\Omega}=\left|f(\theta)\right|^2=\left|f_{\rm ES}(\theta)
 +f_{\rm ET}(\pi-\theta)\right|^2, \label{eq2}
\end{equation}  
where the elastic $\alpha$ transfer amplitude at the angle ($\pi-\theta$) is 
coherently added to the elastic scattering amplitude at $\theta$ in the
center-of-mass (c.m.) frame as shown kinematically in Fig.~\ref{f1}. The two 
amplitudes can be expressed in terms of the partial wave expansion as 
\begin{equation}
 f_{\rm ES}(\theta)=f_{\rm R}(\theta)+\frac{1}{2ik}\sum_\ell (2\ell+1)
 e^{2i\sigma_\ell}\left(S^{\rm ES}_\ell-1\right)P_\ell(\cos\theta), \label{eq3}
\end{equation}
where $f_{\rm R}(\theta)$ and $\sigma_\ell$ are the Rutherford scattering amplitude
and Rutherford phase shift, respectively, which are available in the analytical 
form \cite{Sat83},    
\begin{eqnarray}
 f_{\rm ET}(\theta)&=&\frac{1}{2ik}\sum_\ell(2\ell+1)e^{2i\sigma_\ell}
  S^{\rm ET}_\ell P_\ell(\cos(\pi-\theta)), \nonumber\\
  &=& \frac{1}{2ik}\sum_\ell(2\ell+1)e^{2i\sigma_\ell}S^{\rm ET}_\ell 
  (-1)^\ell P_\ell(\cos\theta). \label{eq4}
\end{eqnarray}
The total elastic amplitude is then obtained as
\begin{equation}
 f(\theta)=f_{\rm R}(\theta)+\frac{1}{2ik}\sum_\ell(2\ell+1)e^{2i\sigma_\ell}
 (S_\ell-1)P_\ell(\cos\theta), \label{eq5}
\end{equation}
\begin{equation}
{\rm where}\ S_\ell=S^{\rm ES}_\ell+(-1)^\ell S^{\rm ET}_\ell. \label{eq6}
\end{equation}
We have thus obtained the total elastic amplitude (\ref{eq6}) in the same 
partial wave expansion as the purely elastic scattering amplitude (\ref{eq3}), 
with an angular-momentum or parity dependent contribution from the elastic 
$\alpha$ transfer added to the elastic scattering $S$-matrix. The interference 
between these two terms in the total $S$-matrix (\ref{eq6}) gives rise naturally 
to an oscillation of elastic cross section at large angles, similar to that caused
by the Mott oscillation observed in the elastic scattering of two identical nuclei, 
like \cc or \oo \cite{Kho16}. We note that the ALAS at backward angles was suggested 
to be due, in general, to some interference that can be generated in the OM calculation 
by adding an angular-momentum or parity dependent term to the total OP 
\cite{Brau82,vOe75}. In particular, the dynamic coupling between the elastic 
scattering and elastic transfer channels was suggested by Frahn and Hussein 
\cite{Frahn80,Frahn80s,Frahn84} to lead to an effective coupling potential 
that depends explicitly on the angular momentum. Moreover, it was also shown 
by these authors \cite{Frahn80s} that the contribution of such dynamic $\ell$-dependent 
coupling potential can be represented by a modified elastic $S$-matrix that contains 
an $\ell$-dependent component like that in Eq.~(\ref{eq6}).  
      
\begin{figure}\vspace*{-1cm}\hspace*{-0.3cm}
\includegraphics[width=0.54\textwidth]{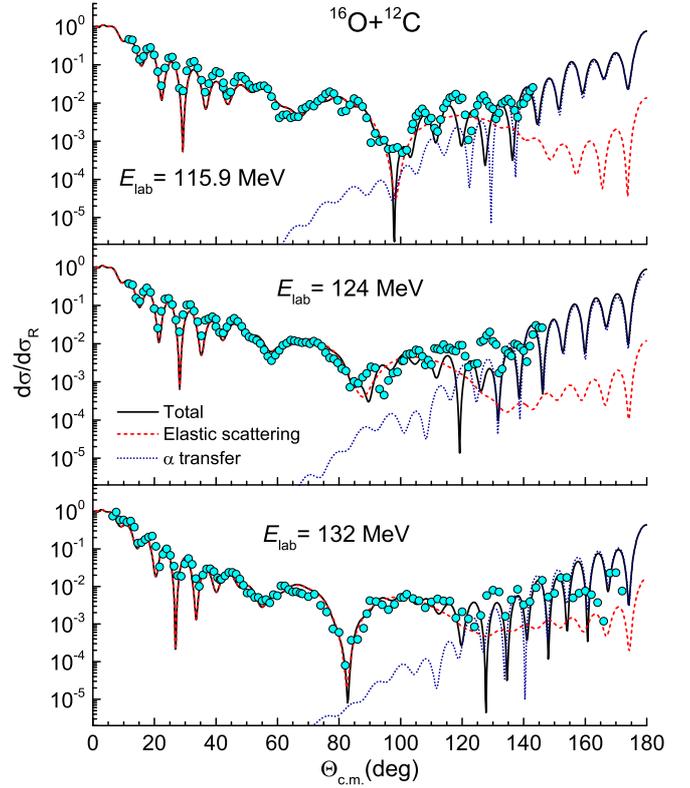}\vspace*{-1cm}
 \caption{CRC description of the elastic \oc data measured at $E_{\rm lab}=115.9$, 
124 MeV \cite{Nico00}, and 132 MeV \cite{Oglo98,Oglo00} using the real folded 
and WS imaginary OP (solid lines). The purely elastic scattering and direct 
elastic $\alpha$ transfer cross sections are shown as the dotted and dashed
lines, respectively. See more details in the text.} \label{f3}
\end{figure}
The total elastic \oc cross sections given by the two-channel CRC calculation 
are compared with elastic \oc data measured at $E_{\rm lab}=115.9$, 
124 MeV \cite{Nico00}, and 132 MeV \cite{Oglo98,Oglo00} in Fig.~\ref{f3}, and 
one can see that the enhanced oscillating elastic cross sections at backward 
angles are due entirely to the elastic $\alpha$ transfer \Atrans process. 
The back coupling of the direct $\alpha$ transfer at backward angles to the 
purely elastic \oc scattering at forward angles turned out to be quite weak, 
and the OP used in the OM calculation discussed in Sec.~\ref{sec2} was used
also in the two-channel CRC calculation. It is remarkable that the good CRC 
description of elastic \oc cross section at backward angles at three energies 
has been reached consistently with the $\alpha$ spectroscopic factor 
$S_\alpha\approx 1.96$, in agreement with those deduced ealier from the DWBA 
and two-channel CRC calculations of elastic \oc scattering at low energies 
\cite{Mor11,Szi02}. Although such a value of the $\alpha$ spectroscopic 
factor is larger than that predicted by the SM calculation \cite{Volya15} or 
4$\alpha$ cluster model \cite{Yama12}, it can be considered as the effective 
$S_\alpha$ that accounts for the elastic $\alpha$ transfer in the two-channel 
approximation. In fact, the comprehensive CRC calculation of elastic 
\oc scattering including explicitly up to 10 reaction channels of both the direct 
and indirect (multistep) $\alpha$ transfers \cite{Phuc18} accounts equally well 
for the elastic data measured at backward angles, using $S_\alpha$ predicted by 
the SM calculation \cite{Volya15}. It is, thus, sufficient to investigate the 
impact by the elastic $\alpha$ transfer on nuclear rainbow based on the results 
of the two-channel CRC calculation only.

\begin{figure}\vspace*{-1cm}\hspace*{-0.5cm}
\includegraphics[width=0.55\textwidth]{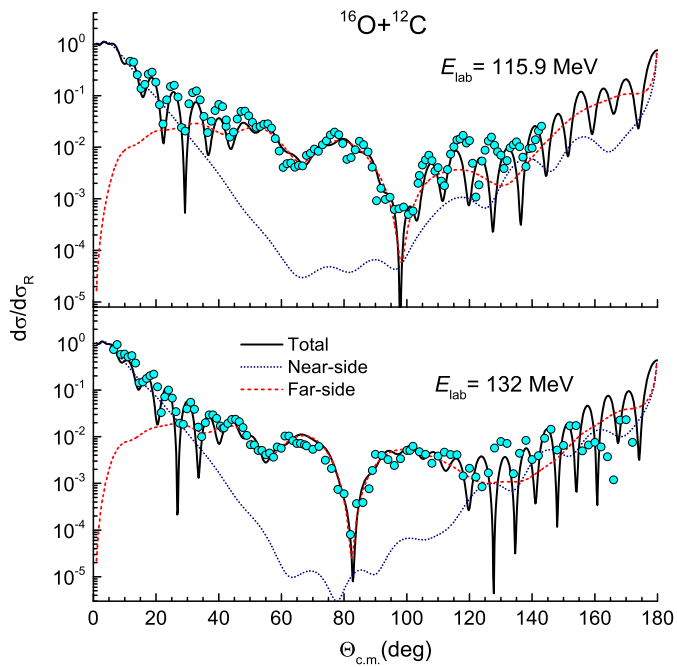}\vspace*{-0.5cm}
 \caption{Two-channel CRC description (solid lines) of elastic \oc data 
measured at $E_{\rm lab}=115.9$ MeV \cite{Nico00} and 132 MeV \cite{Oglo98,Oglo00}. 
The total elastic \oc cross section has been decomposed into the near-side 
(dotted lines) and far-side (dashed lines) contributions (\ref{eq1}) using 
the Fuller method \cite{Ful75}.} \label{f4}
\end{figure}
As discussed above in Sec.~\ref{sec2}, the oscillatory elastic \oc cross 
sections seen at backward angles could be reproduced in the standard OM 
calculation only by using an extremely small diffuseness of the WS imaginary 
OP which enhances the near-side scattering at large angles, and the 
near-far interference there gives rise to a quick oscillation of elastic cross 
section \cite{Phuc18}. Given the strong impact of the elastic $\alpha$ transfer 
shown in Fig.~\ref{f3}, we have decomposed the total elastic amplitude (\ref{eq6}) 
into the near-side and far-side components (\ref{eq1}), and the results 
are shown in Fig.~\ref{f4}. One can see that the elastic $\alpha$ transfer leads 
indeed to the enhanced strength of the near-side scattering at backward angles, 
and the near-far interference results on the oscillating elastic \oc cross 
section. At the given scattering angle $\theta$, the elastic $\alpha$ transfer 
amplitude calculated at the angle ($\pi-\theta$) is added to the purely elastic 
scattering amplitude at $\theta$. Therefore, the enhancement of the total elastic 
\oc cross section at backward angles is in fact caused by the elastic $\alpha$ 
transfer occuring physically at forward angles, at the surface of two colliding 
nuclei. This is a strong indication that the $\alpha+^{12}$C cluster configuration 
is likely formed at the surface of the $^{16}$O nucleus. The same conclusion was 
also drawn from the multichannel CRC analysis of the elastic \oc scattering that 
included the direct as well as indirect $\alpha$ transfer channels \cite{Phuc18}. 
As a consequence, the smooth Airy pattern of nuclear rainbow in the elastic 
\oc scattering at low energies is strongly distorted by the near-far interference 
at backward angles that is due mainly to the elastic $\alpha$ transfer.  

We note that the CRC results shown here were obtained with the WS imaginary 
OP that has a normal diffuseness $a_V\approx 0.5-0.6$ fm. Therefore, the 
abnormal surface absorption of the OP given by an extremely small 
diffuseness of the WS imaginary OP \cite{Nico00} just mimics the strong 
dynamic polarization of the OP by the reaction-channel coupling between 
the purely elastic scattering and elastic $\alpha$ transfer channels.   

\begin{figure}\vspace*{-1cm}\hspace*{-0.5cm}
\includegraphics[width=0.55\textwidth]{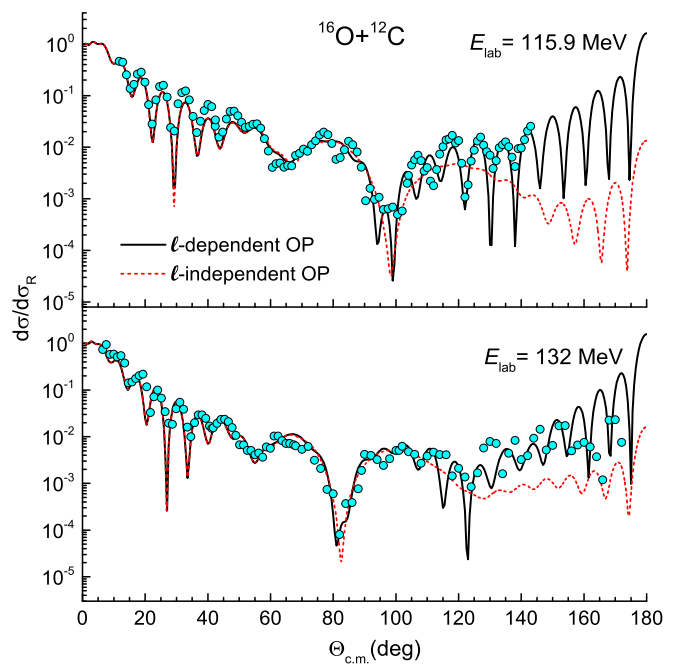}\vspace*{-0.5cm}
 \caption{OM description of the elastic \oc data measured at $E_{\rm lab}=115.9$ 
MeV \cite{Nico00} and 132 MeV \cite{Oglo98,Oglo00}, using the same $\ell$-independent 
OP as in Fig.~\ref{f2} (dashed lines), and that added by an imaginary $\ell$-dependent 
potential (\ref{eq7}) parametrized in the present work (solid lines).} \label{f5}
\end{figure}

It was shown explicitly by Frahn and Hussein 40 years ago \cite{Frahn80} that the 
dynamic coupling between the elastic scattering and elastic transfer channels can 
be effectively taken into account in the OM calculation by adding an $\ell$-dependent 
term to the OP. In fact, the angular-momentum or parity-dependent OP was used 
earlier to account for the oscillating ALAS pattern at backward angles in the elastic 
light HI scattering at low energies \cite{Brau82,vOe75}. To further explore this 
important conclusion we have added to the WS imaginary OP of elastic \oc scattering 
at $E_{\rm lab}=115.9$ and 132 MeV an $\ell$-dependent term $W_\ell(R)$ parametrized 
in the form suggested in Ref.~\cite{Frahn80} 
\begin{equation}
  W_\ell(R)=(-1)^\ell W_s \frac{\exp[-\beta^2(R-R_s)^2]}{\beta R}
	\equiv (-1)^\ell W_{\rm M}(R), \label{eq7}
\end{equation}
where $W_{\rm M}(R)$ is often discussed as the Majorana potential \cite{Phuc19}, 
which accounts for the core-core symmetry of a light HI system like \oc. 
The best OM fit to the elastic \oc data at $E_{\rm lab}=115.9$ and 
132 MeV over the whole angular range has been achieved with $R_s\approx 5.55$ fm, 
so that the $\ell$-dependent contribution (\ref{eq7}) of the OP is peaked at the 
surface. The remaining parameters depend slightly on energy, $W_s\approx 1.834$ MeV 
and $\beta\approx 0.775$ fm at $E_{\rm lab}=115.9$ MeV, and $W_s\approx 1.885$ MeV 
and $\beta\approx 0.806$ fm at $E_{\rm lab}=132$ MeV. The enhanced oscillating 
elastic \oc cross sections at backward angles are now well reproduced by the OM 
calculation using the $\ell$-dependent OP (see Fig.~\ref{f5}), in about the same 
way as the two-channel CRC calculation discussed in Sec.~\ref{sec2}. Thus, the dynamic 
polarization of the OP by the reaction-channel coupling between the purely elastic 
scattering and direct elastic $\alpha$ transfer channels can be effectively accounted 
for by the $\ell$-dependent potential (\ref{eq7}) added to the imaginary OP in the 
standard one-channel OM calculation.  

\begin{figure}\vspace*{-0.5cm}\hspace*{-0.8cm}
\includegraphics[width=0.61\textwidth]{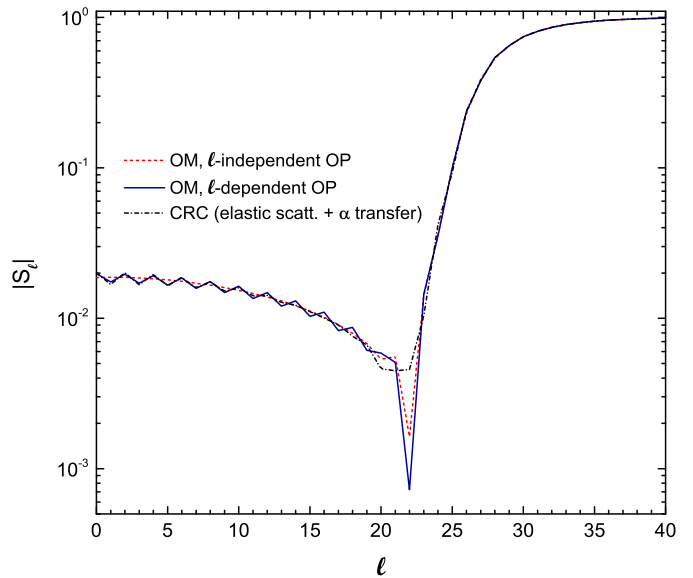}\vspace*{-0.5cm}
 \caption{$S$-matrix elements given by the OM calculation of the elastic \oc scattering 
at 132 MeV using the $\ell$-independent OP (dashed line) and $\ell$-dependent OP 
(solid line), which give the elastic cross sections shown in lower panel of Fig.~\ref{f5}. 
The $S$-matrix given by the two-channel CRC calculation is shown as dash-dotted line.} 
\label{f6}
\end{figure}
To illustrate the dynamic impact of the $\ell$-dependent potential (\ref{eq7}),
we have plotted in Fig.~\ref{f6} the elastic $S$-matrix elements given by both 
the $\ell$-independent and $\ell$-dependent OP's in comparison with the elastic 
$S$-matrix given by the two-channel CRC calculation. The $S$-matrix given by the 
$\ell$-dependent OP turns out to have the same zigzag behavior at small partial 
waves $\ell$ as that of the total elastic $S$-matrix obtained from the two-channel 
CRC calculation. In fact, such an odd-even staggering in the $\ell$ dependence 
of the total elastic $S$-matrix is due to the $\ell$-dependent contribution from 
the elastic $\alpha$ transfer amplitude in Eq.~(\ref{eq6}). The OM results shown 
in Figs.~\ref{f5} and \ref{f6} also indicate that the use of an $\ell$-dependent 
OP might be necessary in the one-channel OM description of the elastic light HI 
scattering at low energies, to effectively account for the dynamic polarization 
of the OP by the nonelastic reaction channels (see more detail in a recent review 
by Mackintosh \cite{Mack19}).
\begin{figure}\vspace*{0cm}\hspace*{-0.3cm}
\includegraphics[width=0.55\textwidth]{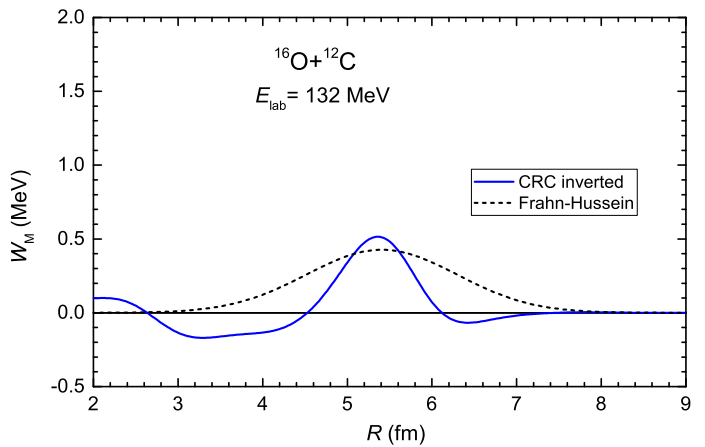}\vspace*{-0.3cm}
 \caption{Imaginary part of the Majorana potential deduced by the IP inversion
method \cite{Phuc19,Mack19} from the $S$ matrix given by the two-channel CRC 
calculation of the elastic \oc scattering at 132 MeV (solid line), in comparison 
with that of the best-fit Majorana potential (dahsed line) parametrized in the 
Gaussian form (\ref{eq7}) suggested by Frahn and Hussein \cite{Frahn80}.} \label{f7}
\end{figure}

We note that the symmetric exchange of the two $^{12}$C cores in the elastic 
\oc scattering has been recently confirmed \cite{Phuc19} to give rise to the parity 
dependence of the OP. Namely, the $S$-matrix generated by the CRC calculation of elastic 
\oc scattering \cite{Phuc18} was used as the input for the iterative-perturbative 
(IP) inversion method \cite{Mack19} to obtain the effective local OP that contains 
both the ($\ell$-independent) Wigner and ($\ell$-dependent) Majorana terms. The results 
of the IP inversion were obtained with high precision and a strong Majorana term 
has been deduced for the total OP of the \oc system, which is a direct estimation 
of the $\ell$-dependent contribution by the core exchange or $\alpha$ transfer. 
It is of interest for the present study to compare the best-fit Majorana potential 
(\ref{eq7}) with the imaginary part of the inverted Majorana potential \cite{Phuc19}, 
and results plotted in Fig.~\ref{f7} show indeed that the dynamic polarization 
of the OP by the core exchange or elastic $\alpha$ transfer can be effectively 
taken into account by the $\ell$-dependent potential in the Gaussian form (\ref{eq7}) 
suggested by Frahn and Hussein \cite{Frahn80}. It is interesting that the results 
of the two independent approaches shown in Fig.~\ref{f7} suggest clearly that 
the elastic $\alpha$ transfer occurs mainly at $R\approx 5.4$ fm, i.e., at the 
surface of the \oc system.

\section{Summary}
The effect of the elastic $\alpha$ transfer on the nuclear rainbow pattern in 
the \oc system has been investigated in the two-channel CRC analysis of elastic
\oc scattering, taking explicitly into account the coupling between the purely 
elastic scattering and direct elastic $\alpha$ transfer channels. The elastic 
$\alpha$ transfer was found to enhance strongly the near-side scattering cross 
section at backward angles, giving rise to an oscillating distortion of the 
smooth Airy structure of nuclear rainbow. 

The one-channel OM was shown to give about the same good description of elastic 
\oc data over the whole angular range as that given by the CRC calculation  
if the dynamic polarization of the OP by the coupling between the elastic 
scattering and $\alpha$ transfer channels is effectively taken into account by 
an $\ell$-dependent Majorana potential in the Gaussian form (\ref{eq7}) suggested 
40 years ago by Frahn and Hussein \cite{Frahn80}. This result validates the use 
of the $\ell$-dependent OP in the extensive OM analysis of elastic scattering data 
measured at low energies for those light HI systems that have the core-core 
symmetry, where the elastic nucleon- or cluster transfer can be significant. 

\section*{Acknowledgments}
With the results presented in this work, we would like to appreciate the 
essential and very valuable contribution to the nuclear scattering theory 
by Mahir Hussein, especially, the modeling of the dynamic polarization of 
the nuclear OP by nonelastic channels. The present research has been supported, 
in part, by the National Foundation for Science and Technology Development 
(NAFOSTED Project No. 103.04-2016.35).


\begin{thebibliography}{99}
\bibitem{Sa79} G.R. Satchler and W.G. Love, Phys. Rep. {\bf 55}, 183 (1979).
\bibitem{Bra97} M.E. Brandan and G.R. Satchler, Phys. Rep. {\bf 285}, 143 (1997).
\bibitem{Kho07r} D.T. Khoa, W. von Oertzen, H.G. Bohlen, and S. Ohkubo,
 J. Phys. G {\bf 34}, R111 (2007).
\bibitem{Hus84}M.S. Hussein and K.W. McVoy, Prog. Part. Nucl. Phys. {\bf 12}, 103 (1984).
\bibitem{Bra96} M.E. Brandan, M.S. Hussein, K.W. McVoy, and G.R. Satchler, 
	{\it Comments on nuclear and particle physics}, Vol.~22 (Gordon and Breach, 
	New York, 1996), p.~77. 
\bibitem{Fri88} S.H. Fricke, M.E. Brandan, and K.W. McVoy,
	Phys. Rev. C {\bf 38}, 682 (1988).
\bibitem{Bri77} D.M. Brink and N. Takigawa, Nucl. Phys. A {\bf 279}, 159 (1977).
\bibitem{Row77} N. Rowley, H. Doubre, and C. Marty, Phys. Lett. B {\bf 69}, 
 147 (1977).
\bibitem{Ful75} R.C. Fuller, Phys. Rev. C {\bf 12}, 1561 (1975).
\bibitem{Kho16} D.T. Khoa, N.H. Phuc, D.T. Loan, and B.M. Loc,
	Phys. Rev. C {\bf 94}, 034612 (2016).
\bibitem{Sti89} E. Stiliaris, H. G. Bohlen, P. Fr\"obrich, B. Gebauer, D. Kolbert, 
 W. von Oertzen, M. Wilpert, and Th. Wilpert, Phys. Lett. B {\bf 223}, 291 (1989).
\bibitem{Bra91} M.E. Brandan and G.R. Satchler, Phys. Lett. B {\bf 256}, 311 (1991).
\bibitem{Rou85} P. Roussel, N. Alamanos, F. Auger, J. Barrette, B. Berthier, 
 B. Fernandez, L. Papineau, H. Doubre, and W. Mittig, Phys. Rev. Lett. {\bf 54}, 1779 (1985).
\bibitem{Bra86} M.E. Brandan, A. Menchaca-Rocha, M. Buenerd, J. Chauvin, 
 P. De Saintignon, G. Duhamel, D. Lebrum, P. Martin, G. Perrin, and J.Y. Hostachy, 
 Phys. Rev. C {\bf 34}, 1484 (1986). 
\bibitem{Vil89} A.C.C. Villari, A. L\'{e}pine-Szily, R.L. Filho, O.P. Filho, M.M. Obuti, 
 J.M. Oliveira Jr, and N. Added, Nucl. Phys. A  {\bf 501}, 605 (1989).
\bibitem{Oglo98} A.A. Ogloblin, D.T. Khoa, Y. Kond\=o, Yu.A. Glukhov, 
 A.S. Dem’yanova, M.V. Rozhkov, G.R. Satchler, and S.A. Goncharov, 
 Phys. Rev. C {\bf 57}, 1797 (1998). 
\bibitem{Oglo00} A.A. Ogloblin, Yu.A. Glukhov, W.H. Trzaska, A.S. Demyanova, 
	S.A. Goncharov, R. Julin, S.V. Klebnikov, M. Mutterer, M.V. Rozhkov, 
	V.P. Rudakov, G.P. Tiorin, D.T. Khoa, and G.R. Satchler, 
	Phys. Rev. C {\bf 62}, 044601 (2000).
\bibitem{Glu01-181} Yu. A. Glukhov, S.A. Goncharov, A.S. Demyanova, A.A. Ogloblin, 
 M.V. Rozhkov, V.P. Rudakov, and V. Trashka, 
 Izv. Ross. Akad. Nauk, Ser. Fiz. {\bf 65}, 647 (2001).
\bibitem{Glu07} Yu. A. Glukhov, V. P. Rudakov, K. P. Artemov, A. S. Demyanova, 
 A. A. Ogloblin, S. A. Goncharov, and A. Izadpanakh, Phys. At. Nucl. {\bf 70}, 1 (2007). 
\bibitem{Nico00} M.P. Nicoli, F. Haas, R.M. Freeman, S. Szilner, Z. Basrak, A. Morsad, 
 G.R. Satchler, and M. E. Brandan, Phys. Rev. C  {\bf 61}, 034609 (2000).
\bibitem{Bra01} M. E. Brandan, A. Menchaca-Rocha, L. Trache, H.L. Clark,
	A. Azhari, C. A. Gagliardi, Y.-W. Lui, R. E. Tribble, R. L. Varner,
	J. R. Beene, and G. R. Satchler, Nucl. Phys. A {\bf 688}, 659 (2001).
\bibitem{Kho94} D.T. Khoa, W. von Oertzen, and H.G. Bohlen, 
 Phys. Rev. C {\bf 49}, 1652 (1994).
\bibitem{Kho97} D.T. Khoa, G.R. Satchler, and W. von Oertzen,
 Phys. Rev. C {\bf 56}, 954 (1997).
\bibitem{Bra88} M.E. Brandan and G.R. Satchler, Nucl. Phys. A {\bf 487}, 477 (1988).
\bibitem{Kho00a} D.T. Khoa, W. von Oertzen, H.G. Bohlen, and F. Nuoffer, 
	Nucl. Phys. A {\bf 672}, 387 (2000).
\bibitem{Mich01} F. Michel, G. Reidemeister, and S. Ohkubo, 
 Phys. Rev. C {\bf 63}, 034620 (2001).
\bibitem{Ohku14-1} S. Ohkubo and Y. Hirabayashi, Phys. Rev. C {\bf 89}, 051601(R) (2014).
\bibitem{Brau82} P. Braun-Munziger and J. Barette, Phys. Rep. {\bf 87}, 209 (1982).
\bibitem{Frahn80} W.E. Frahn and M.S. Hussein, Phys. Lett. {\bf 90B}, 358 (1980).
\bibitem{Frahn80s} W.E. Frahn and M.S. Hussein, Nucl. Phys. A {\bf 346}, 237 (1980).
\bibitem{Frahn84} W.E. Frahn, {\it Treaties on Heavy-Ion Science} vol.~1, p.~135, 
 ed. D.A. Bromley (Plenum Press, New York, 1984).
\bibitem{Phuc18} N.T.T. Phuc, N.H. Phuc, and D.T. Khoa, 
 Phys. Rev. C {\bf 98}, 024613 (2018).
\bibitem{vOe75} W. von Oertzen and H.G. Bohlen, Phys. Rep. {\bf 19 C}, 1 (1975).
\bibitem{Mor11} M.C. Morais and R. Lichtenth\"{a}ler, Nucl. Phys. A {\bf 857}, 1 (2011).
\bibitem{Hama11} Sh. Hamada, N. Burtebayev, K.A. Gridnev, and N. Amangeldi, 
 Nucl. Phys. A {\bf 859}, 29 (2011). 
\bibitem{Szi02} S. Szilner, W. von Oertzen, Z. Basrak, F. Haas, and M. Milin,
	Eur. Phys. J. A {\bf 13}, 273 (2002). 
\bibitem{Rud10} A.T. Rudchik {\it et al.}, Eur. Phys. J. A {\bf 44}, 221 (2010).
\bibitem{Raynal} J. Raynal, {\it Computing as a Language of Physics}
 (IAEA, Vienna, 1972) p.~75;  J. Raynal, coupled-channel code ECIS97 (unpublished).
\bibitem{Volya15} A. Volya and Y. M. Tchuvilsky, Phys. Rev. C {\bf 91}, 044319 (2015).
\bibitem{Yama12} T. Yamada, Y. Funaki, T. Myo, H. Horiuchi, K. Ikeda, G. R\"opke,
 P. Schuck, and A. Tohsaki, Phys. Rev. C {\bf 85}, 034315 (2012).
\bibitem{Tho88} I.J. Thompson, Comput. Phys. Rep. {\bf 7}, 167 (1988); 
 http://www.fresco.org.uk.
\bibitem{Sat83} G.R. Satchler, {\it Direct Nuclear Reactions} (Clarendon, Oxford, 1983).
\bibitem{Tho09} I.J. Thompson and F.M. Nunes, {\it Nuclear Reactions for Astrophysics} 
(Cambridge University Press, Cambridge, UK, 2009).
\bibitem{Phuc19} N.T.T. Phuc, R.S. Mackintosh, N.H. Phuc, and D.T. Khoa, 
 Phys. Rev. C {\bf 100}, 054615 (2019).
\bibitem{Mack19} R.S. Mackintosh, Eur. Phys. J. A {\bf 55}: 147 (2019). 
\end{thebibliography}
\end{document}